\documentstyle[12pt]{article}
\newcommand{\be}{\begin{eqnarray}}
\newcommand{\ee}{\end{eqnarray}}
\def\cc{\hbox{C\kern-0.55em\raise0.4ex\hbox{$\scriptstyle |$}}}
\def\rr{\hbox{{\sf R}\kern-0.50em\raise0.0ex\hbox{\sf R}}}
\begin{document}
\sl

\begin{titlepage}
\rightline{FTUV/93-11}
\begin{center}
\vspace*{1.5cm}
{\bf STAR PRODUCTS AND DEFORMED OSCILLATORS}\footnote
{Supported by DGICYT, Spain}
\vskip 0.5 cm
{\bf Demosthenes Ellinas}\\
Departamento de Fisica Teorica and IFIC\\
Centro Mixto Universidad de Valencia-CSIC \\
Valencia Spain E-46100 Burjasot\\
ellinas@evalvx.ific.uv.es
\end{center}
\vskip 2.0 cm
\begin{abstract}
{\sl A star-product formalism describing deformations of the standard
quantum mechanical harmonic
oscillator is introduced. A number of existing generalized oscillators
occur as particular choises of star-products between the elements of the
ordinary oscillator algebra. Star dynamics and coherent states are introduced
and studied.}
\end{abstract}
\end{titlepage}

The recent activity in the theory of quantum groups [1,2,3] has invigorated
the interest in the central issue  of quantum theory, namely quantization.
Instrumental in the studies of quantum groups is the notion of deformation.
Quantization of a group is understood as deformation of the algebra of
functions of the group manifold and as deformation of the respective
enveloping algebra of generators, both using a deformation parameter $q$,
to which in principle should be attributed a physical meaning. On the
other hand, quantization of classical mechanics also can be understood, outside
the convential Hilbert space - with operators framework, as a deformation
of the commuting algebra of functions defined on the classical phase space,
with deformation parameter the Planck constant. This deformation
quantization scheme is usually called star-product quantization [4,5]
and its roots are dated back to the early days of quantum mechanics when
it was known as Weyl-Wigner quantization [6,7,8].

In this paper we will attempt to describe the algebra quantization along
the lines of the star-quantization method, using the $q$-deformed
oscillator as a paradigm. Our "classical" algebra of functions will be
the noncommutative enveloping algebra of creation and annihilation operators
$a^\pm$, of the standard oscillator algebra of quantum mechanics $h_4$.
Its deformation will proceed with the introduction of two different
star-products defined between elements of the oscillator algebra. These
products are operator valued and depend by construction only upon the number
operator $N$, and a deformation parameter $q$. Utilizing these star-products
we deform the oscillator algebra in a manner which affects only the
commutation relation between the $a^+$ and $a^-$ and keeps unchanged
all the rest, this algebra we call $*-h_4$. Then we proceed by showing
the connection between the present star-product method of deforming
the harmonic oscillator and the customary prescription of the so called
$q$-deformed oscillator. The connection is established by means of the
quantizing mappings relating deformed with undeformed generators.
As is explained later in the text the introduction of two different
star-products is necessary in order to obtain the $q$-oscillator after a
particular choise of products. It is also shown how other generalized
oscillators  are reached by making other choises of star-products.
Finally, dynamics and coherent states pertinent to $*-h_4$ are introduced
and studied to some extend.

The star-quantization program constitutes an autonomous alternative to
quantization [4,5]. The basic idea is to start with the commutative algebra
of functions in the oscillator phase space with pointwise multiplication
which is also
a Lie algebra under the Poisson bracket. This algebra is now deformed with
deformation parameter the $\hbar$. Let $(q,p)\equiv(z_1,z_2)\equiv(z)$ then
for $f(z)$ and $g(z)$ functions of the phase space the Poisson bracket is
$(\partial_z\equiv\frac{\partial}{\partial z})$
\be
\biggl\{ f{(z)},g{(z)}\biggr\}_{PB}=f{(z)}\Lambda g{(z)}= f{(z)}
(\stackrel{\leftarrow}{\partial}_{z_1}\stackrel{\leftarrow}{\partial}
_{z_2})\left(\begin{array}{cc}0 &
-1\\1&0\end{array}\right)\left(\begin{array}{c
}
\stackrel{\rightarrow}{\partial}_{z_1}\\ \stackrel{\rightarrow}{\partial}_{z_2}
\end{array}\right) g{(z)}\ .
\ee
If for example we restrict ourselves to functions which are up to quadratic
polynomials we obtain the $s\ell_2$ algebra or if we allow for polynomials
of infinite-degree (series) we obtain a infinite dimensional Lie algebra etc.
We note also that the Poisson bracket is skew-symmetric, $\{ f,g\}=-\{
g,f\}$, satisfies the Jacobi identity $\{ f,\{ g,h\}\}+\{g,\{ g,f\}\}+\{
h,\{f,g\}\}=0$ and obeys the Leibnitz rule, $\{ f,gh\}=\{ f,g\}+g\{ f,h\}$.
Now the deformation of such an algebra is obtained by first imposeing
a non-commutative $*$-product between the functions
\be
f(z)*g(z):=f(z)e^{\frac{i\hbar}{2}\Lambda}g(z)\ ,
\ee
of course
\be
f*g=f\cdot g+\frac{i\hbar}{2}\biggl\{ f, g\biggr\}+{\cal O}(\hbar^2)\ .
\ee
Then the so-called Moyal bracket is defined by
\be
f{(z)}*g{(z)}-g{(z)}*f(z)=i \hbar\biggl\{ f{(z)},g{(z)}\biggr\}+{\cal
O}(\hbar^2
)
=:\biggl\{f{(z)},g{(z)}\biggr\}_*
\ee
and is equivalent to the commutator of quantum mechanics, ordered such that
the position operator is placed to the left of the momentum operator.
It must be evident now
that the philosophy of the $*$-quatization is entirely opposite to the
quantum-mechanics-with-Hilbert-space method of ordinary quantization.
The emphasis
here has been shifted from the objects (operators) to the rule of
composition (*-product) between the objects (classical functions).
This approach to quantization is in fact similar to the Weyl-Wigner
formalism of quantum mechanics (see e.g. Ref. 6,7,8), and to the Berezin
type of quantization through coherent states by the so-called symbols of
operators [9].

The Moyal bracket constitutes a non-trivial deformation of the Poisson
bracket and is essential that it involves derivatives of infinite order,
since according to Kirillov [10] any deformation  depending on a finite
number of jets of functions is trivial at least locally (Darboux theorem).
Moreover the Moyal bracket is the unique, up to isomorphism, non-trivial
deformation of the Poisson bracket [10, 11, 12]. (These and subsequent
statement
s
concerning the $*$-products and Moyal brackets are valid for any number
of degrees of freedom, however here we state them for only one degree of
freedom) .

Two important tools of this formalism are first the
*-exponential of a function,
\be
e_*^{1/i\hbar f}=\sum\limits^\infty_
{n=0}\frac{1}{n!}\biggl(\frac{1}{i\hbar}\biggr)^n
(f*)^n
\ee
where $(f*)^n:=f*\cdots *f$ ($n$-times). If $H$ is the classical Hamiltonian
then
\be
e_*^{t/i\hbar H}=\sum\limits^\infty_{n=0}
\frac{1}{n!}\biggl(\frac{t}{i\hbar}\biggr)^n(H*)^n
\ee
and the time development of any function of the phase space is be given by,
\be
f(t)=e_*^{-(\frac{t}{i\hbar}H)}*f(0)*e_*^{t/i\hbar H}\ ,
\ee
then $f(t)$ obeys
\be
i\hbar \frac{df}{dt}=f*H-H*f=\biggl\{ f,H\biggr\}_*
\ee
which is the full quantum equation of motion of this formalism.

The second tool is the star Backer-Champbel-Hausdorff, *-BCH, formula;
the symbols $a(z),\ b(z)$ and $c(z)$
of three operators $A,B$ and $C$ for which the BCH formula is valid, namely
\be
e^Ae^B=e^C
\ee
with
\be
C=A+B+L_2+L_3+\cdots+L_n+\cdots
\ee
and
\be
L_2&=&\frac{1}{2}[A,B]\nonumber\\
L_3&=&\frac{1}{12}[A,[A,B]]+\frac{1}{12}[[A,B],B]\hspace{1.0cm}{\rm
etc.}\nonumb
er
\ee
satisfy the $*$-BCH formula, namely:
\be
e^a_**e^b_*=e^c_*
\ee
where
\be
c=a+b+\ell_2+\ell_3+\cdots+\ell_n+\cdots
\ee
with
\be
\ell_2&=&\frac{1}{2}\biggl\{ a,b\biggr\}_*\nonumber\\
\ell_3&=&\frac{1}{12}\biggl\{ a,\biggl\{ a,b\biggr\}_*\biggr\}_*+\frac{1}{12}
\biggl\{\biggl\{ a,b\biggr\}_*,b\biggr\}_*\hspace{1.0cm}{\rm etc.}\nonumber
\ee
Let us finally note that if $\alpha=\frac{1}{\sqrt{2}}(q+ip)$ and
$\bar{\alpha}=\frac{1}{\sqrt{2}}(q-ip)$ then the fundamental Poisson bracket of
the
classical oscillator
\be
\biggl\{\alpha,\bar{\alpha}\biggr\}_{PB}=1,
\ee
is in the $*$-product formalism replaced by the Moyal bracket,
\be
\biggl\{\alpha,\bar{\alpha}\biggr\}_*=1,
\ee
which now describes the quantum-mechanical oscillator.
\vskip 0.5 cm
  Let us now come to the $q$-deformed oscillator. The $q$-deformation of the
qua
ntum-mechanical
harmonic oscillator as it is known in the literature [13, 14, 15]
is along the lines of the quantization of the classical
oscillator by means of the ordinary
method of operators and Hilbert spaces.
Since we expect that conceptual as well as technical profit can be
made by recasting the $q$-deformation into some alternative formalism of
a $q$-star-product type, we will proceed as the following diagram indicates:

\pagebreak

\setlength{\unitlength}{1mm}
\begin{picture}(150,220)(10,20)
\put(130,30){\makebox(0,0){\large The present work}}
\put(100,35){\framebox(65,20){\shortstack{{\large $q$-Star-product}\\
{\large deformation e.g. $q$-oscillator}}}}
\put(0,70){\framebox(50,20){\shortstack{\large Quantum Mechanics\\ \large e.g
Oscillator}}}
\put(100,85){\framebox(60,30){\shortstack{\large Quantum Groups\\
\large Operator-Hilbert-space\\ \large deformation}}}
\put(100,135){\framebox(50,20){\shortstack{\large $\hbar$-Star-product\\
\large quantization}}}
\put(0,170){\framebox(50,20){\shortstack{\large Classical Mechanics\\
\large e.g Oscillator}}}
\put(100,185){\framebox(60,30){\shortstack{\large Quantum Mechanics\\
\large Operator-Hilbert-space\\ \large quantization}}}
\put(10,220){\makebox(0,0){\large  }}
\put(55,75){\vector(3,-2){40}}
\put(55,85){\vector(3,1){40}}
\put(55,175){\vector(3,-2){40}}
\put(55,185){\vector(3,1){40}}
\end{picture}

\pagebreak

\noindent To this end let us consider the harmonic oscillator algebra
denoted by $h_4$ [16]:
\be
& &\ [a^-,a^+]={\bf 1}\\
\nonumber\\
& &\ [N,a^\pm]=\pm a^\pm\\
\nonumber\\
& &\ [{\bf 1},\ {\rm everything}]=0 .
\ee
Let us endowed this algebra with two $*$-products: the right-star-product
$*_{R}$,
\be
g_1*_Rg_2\equiv g_1\cdot R(N)\cdot g_2\ ,
\ee
and the left-star-product $*_L$,
\be
g_1*_Lg_2=g_1\cdot L(N)\cdot g_2
\ee
for any $g_1,g_2\epsilon h_4$. The $R(N)$ and $L(N)$ are smooth functions of
the
number operator $N$, which depend also on the deformation parameter
$q=e^\gamma,
\
\gamma\in\rr$\ or $\cc$\ \, in such a way that when $q\rightarrow 1$ then $R(N)
\rightarrow{\bf 1}$ and $L(N)\rightarrow{\bf 1}$ (simultaneously).
Each of these $*$-products are non-commutative and for any $g_1,g_2,
g_3\epsilon h_4$
which satisfy the Jacobi identity is valid that
\be
g_1*_Rg_2*_Rg_3+cyclic\ permutation=0
\ee
and
\be
g_1*_Lg_2*_Lg_3+cyclic\ permutation=0\ .
\ee
Let us now define the corresponding Moyal bracket of our case, in terms of
$*_R$ and $*_L$. It is:
\be
\ [g_1,g_2]_{*_{RL}}&\equiv&g_1*_Rg_2-g_2*_Lg_1\\
\nonumber\\
&=& g_1Rg_2-g_2Lg_1\ .
\ee
We will also need the opposite bracket, namely:
\be
\ [g_1,g_2]_{*_{LR}}\equiv g_1*_Lg_2-g_2*g_1=g_1Lg_2-g_2Rg_1\ .
\ee
Then the antisymmetry can be stated as
\be
\ [g_1,g_2]_{*_{RL}}=-[g_2,g_1]_{*_{LR}}\ .
\ee
We now  proceed by defining the $q$-deformed oscillator in the present
$*$-produ
ct approach. The
$q$-deform harmonic oscillator algebra $*-h_4$ is defined to be:
\be
& &\ [a^-,a^+]_{*_{RL}}={\bf 1}\\
\nonumber\\
& &\ [N,a^\pm]=\pm a^\pm\\
\nonumber\\
& &\ [{\bf 1},\ {\rm everything}]=0 .
\ee
The generalized $q$-Jacobi identity valid for that algebra reads [17]
\be
\ [a^-,[a^+,N]]_{*_{RL}}+[a^+,[N,a^-]]_{*_{LR}}+[N,[a^-,a^+]_{*_{RL}}]=0
\ee
or
\be
\ [a^-,[a^+,N]]_{*_{LR}}+[a^+,[N,a^-]]_{*_{RL}}+[N,[a^-,a^+]_{*_{RL}}]=0 .
\ee
Any choice of $*_R$ and $*_L$ products will induce a choise of the Moyal
bracket
and this will determine in turn
a different $q$-oscillator. To come in contact with
quantum groups (this is  abuse of language, $q$-oscillator have no satisfactory
co-product), we take the $R(N)$ and $L(N)$ functions to be as follows,
\be
R(N)\equiv \frac{S(N)}{N}\cdot\frac{1}{S(N)-S(N-1)}
\ee
and
\be
L(N)\equiv\frac{S(N+1)}{N+1}\cdot\frac{1}{S(N+2)-S(N+1)},
\ee
where $S(N)$ parametrizes the products and is a function of $N$ and of the
deformation parameter to be specified later. Let us now see how the
$q$-deformed ocillator algebra of eqs. (26-28) includes different known
deformed
oscillators, by merely choosing different parametrizing functions $S(N)$ which
i
n turn
choose $*_R,\ *_L$ and Moyal brackets. We first elaborate on (26) using
the  identity $f(N)a^\pm=a^\pm f(N\pm 1)$;
\be
a^-*_Ra^+-a^+*_La^-={\bf 1}
\ee
means
\be
a^-R(N)a^+-a^+L(N)a^-={\bf 1}\ ,
\ee
or by virtue of (31-32),
\be
a^-\biggl\{\frac{S(N)}{N}\cdot\frac{1}{S(N)-S(N-1)}\biggr\}a^+-a^+\biggl\{\frac
{S(N+1)}{N+1}\cdot\frac{1}{S(N+2)-S(N+1)}\biggr\} a^-={\bf 1}
\ee
or
\be
\biggl\{\frac{1}{S(N+1)-S(N)}\biggr\}\cdot\biggl\{
a^-\frac{S(N)}{N}a^+-a^+\frac
{S(N+1)}{N+1}a^-\biggr\}={\bf 1}\ ,
\ee
which becomes,
\be
a^-\sqrt{\frac{S(N)}{N}}\sqrt{\frac{S(N)}{N}}a^+-a^+\sqrt{\frac{S(N+1)}
{N+1}}\sqrt{\frac{S(N+1)}{N+1}}a^-=S(N+1)-(N)\ .
\ee
After defining $a^\pm_q$, by means of the deforming mappings
\be
a^-_q\equiv a^-\sqrt{\frac{S(N)}{N}}=\sqrt{\frac{S(N+1)}{N+1}}a^-
\ee
and
\be
a^+_q\equiv a^+\sqrt{\frac{S(N+1)}{N+1}}=\sqrt{\frac{S(N)}{N}}a^+\ ,
\ee
eq. (37) can be rewritten in the form
\be
a^-_qa^+_q-a^+_qa^+_q=S(N+1)-S(N)\ .
\ee
Next we elaborate on (27);
\be
Na^+-a^+N=a^+\ ,
\ee
which can be written as
\be
N\sqrt{\frac{N}{S(N)}}\sqrt{\frac{S(N)}{N}}a^+-a^+\sqrt{\frac{N+1}{S(N+1)}}
\sqrt{\frac{S(N+1)}{N+1}}N=a^+\ ,
\ee
or
\be
N\sqrt{\frac{S(N)}{N}}a^+-a^+\sqrt{\frac{S(N+1)}{N+1}}N=\sqrt{\frac{S(N)}{N}}
a^+\ ,
\ee
or with the definitions issued by (38-39)
\be
Na^+_q-a^+_qN=a^+_q\ ,
\ee
and similarly from $Na^--a^-N=-a^-$ we obtain
\be
Na^-_q-a^-_qN=-a^-_q
\ee
by virtue of the identifications in (38-39). Finally from (28) we get
easily that the unit operator commutes with all the $N,a^+_q$ and $a^-_q$.

Summarizing the results derived for  the deformed generators we have
\be
& &\ [a^-_q,a^+_q]=S(N+1)-S(N)\\
\nonumber\\
& &\ [N,a^\pm_q]=\pm a^\pm_q\\
\nonumber\\
& &\ [{\bf 1},\ {\rm everything}]=0\ .
\ee
One should now contrast eqs. (26-28) with eqs. (46-48); they describe the
same algebra in two fundamentally different ways. The
former is the
*-product road to deformation, the latter is the deformed-operators type
of description of the algebra deformation concept. The bridge between
them is of course the deforming maps of (38-39); they trade the deformed
product (26-28) for deformed generators or the opposite.

If we now select the parametrizing function[18] to be:
\be
S(N)=[N]=\frac{q^N-q^{-N}}{q-q^{-1}}=\frac{\sinh\gamma N}{\sinh\gamma}\ ,
\ee
we obtain from (46),
\be
a^-_qa^+_q-a^+_qa^-_q=[N+1]-[N]
\ee
and from (38,39)[19]
\be
& &a^+_q=a^+\sqrt{\frac{[N+1]}{N+1}}=\sqrt{\frac{[N]}{N}}a^+\ ,\\
\nonumber\\
& &a^-_q=a^-\sqrt{\frac{[N]}{N}}=\sqrt{\frac{[N+1]}{N+1}}a^-\ .
\ee
Equations (50-52) give evidently the $q$-deformed oscillator [13, 14, 15],
or in its original quommutator form
\be
a^-_qa^+_q-qa^+_qa^-_q=q^{-N}\ .
\ee
For the choice (49) the $*_R$ and $*_L$ products are of course in view of
(31-32) given by
\be
*_R=\frac{[N]}{N}\cdot\frac{1}{[N]-[N-1]}
\ee
and
\be
*_L=\frac{[N+1]}{N+1}\cdot\frac{1}{[N+2]-[N+1]}\ .
\ee
We notice however that this scheme includes also other deformed oscillators
with different star-products induced by other choises of the parametrizing
function $S(N)$. The following list indicates some  alternatives:
\vskip 1.0 cm
\begin{tabular}{lll}
$S(N)$ parametrizing *_{R,L} &\hspace{1.0cm} & *-h_4 \hspace{0.3cm} algebra\\
\hline
$N$ & & quantum harmonic oscillator\\ \\
$\frac{q^N-q^{-N}}{q-q^{-1}}$ & & $q$-deformed oscillator [13, 14, 15]\\ \\
$\frac{q^N-1}{q-1}$ & & asymmetric deformed oscillator [20]\\ \\
$\frac{q^N-p^{-N}}{q-p^{-1}}$ & & two-parameter deformed oscillator [21, 22]\\
\
\
$N(p+1-N)$ & & para-fermionic oscillator [23]\\ \\
$\frac{\sin h(\tau N)\sin h(\tau(p+1)-N))}{\sin h^2(\tau)}$ & &
deformed para-fermionic oscillator [24]\\ \\
$N^K$ & & power deformed oscillator [25]\\ \\
$\frac{sn(\tau N)}{sn(\tau)}$ & & elliptic deformed oscillator [25]
\end{tabular}
\vskip 0.5cm

Let us add that this list of alternatives is far from being exausted.
The $su(2)$ algebra, in accordance to the spirit of the present work,
can  also be  derived by means of deformed star-products defined between
the elements of the standard oscillator algebra, and similarly the relation
among other algebras can also be casted in the deformation language as
will be shown elsewere.

  We will  now turn to the study of  the generalized dynamics of the
$q$-oscillator.
We postulate, by
analogy with the $*$-product case as described earlier,
 the following Heisenberg equation
of motion[26, 27]
\be
i\hbar\dot{A}=[A,H]_{*_{RL}}
\ee
This is satisfied by the formal solution
\be
A(t)=e^{\frac{i}{\hbar}HL}A(0)e^{-\frac{it}{\hbar}RH}
\ee
as is easily verified.

The appearance of the $*_{RL}$-commutator in (56) implies that the dynamics is
not canonical i.e the Hamiltonian in general is not $*_{RL}$-commutating with
itself, therefore we have no conservation of energy. Moreover such dynamics is
n
ot an
automorphism for the fundamental commutation relations of the $q$-oscillator
alg
ebra.
These imply a non-canonical dynamics for the $q$-deformed oscillator which
physically accounts for e.g. dissipation, constraints, or non-linear
self interactions.

  To extent our considerations relating the star-products with the algebra
deformation,
we now come to study generalized $*_{RL}$-coherent states pertinent
to the algebra of eq. (26).
To this end we introduce a deformed exponential function in
analogy to the $*$-product formalism of the classical case above.

Let
\be
e^x_{*_{RL}}&\equiv &{\bf 1}_{*_{RL}}+\frac{1}{1!}x+\frac{1}{2!}x\Box
x+\frac{1}
{3!}
x\Box x\Box x+\cdots\nonumber\\
&\equiv&\sum\limits^\infty_{n=0}\frac{1}{n!}(x\Box)^n
\ee
where
\be
{\bf 1}_{*_{RL}}\equiv\biggl(\frac{R+L}{2}\biggr)^{-1}\ ,\nonumber
\ee
and $\Box$ is the abbreviation (not a new product!),
$\Box\equiv\frac{1}{2}(*_R+*_L)$ i.e
\be
x\Box x=\frac{1}{2}x(*_R+*_L)x=\frac{1}{2}(xRx+xLx)\ .
\ee
Obviously $(x\Box)^2\equiv x\Box x\ ^{\mbox{$\displaystyle\longrightarrow$}}
_{q\rightarrow 1}x^2$ and also in the same zero deformation limit,
\be
\left.\begin{array}{cc}\stackrel{q\rightarrow 1}{R\rightarrow{\bf 1}}\\
L\rightarrow{\bf 1}\end{array}\right.\hspace{1.5cm}{\rm and}\hspace{1.5cm}
e^x_{*_{RL}}\stackrel{q\rightarrow 1}{\longrightarrow} e^x .\nonumber
\ee
Two alternative forms of the $*_{RL}$-exponential  function given in terms of
the ordinary exponential
are the following ($x$ any operator),
\be e^x_{*_{RL}}=\biggl(\frac{R+L}{2}\biggr)^{-1}e^{(\frac{R+L}{2})x}
\ee
and
\be
e^x_{*_{RL}}=e^{x(\frac{R+L}{2})}\biggl(\frac{R+L}{2}\biggr)^{-1}\ .
\ee
{}From above we deduce the following composition of the $*_{RL}$-exponentials
for any two commuting elements $x,y$ (with $N$ as well):
\be
e^x_{*_{RL}}\Box e^y_{*_{RL}}=e^{x+y}_{*_{RL}}\ .
\ee
Also for any two non-commuting elements $x,y$ for which the ordinary
BCH formula is valid, i.e
$e^xe^y=e^c$  etc. see eqs. (9-10), then the $*_{RL}$-BCH formula is also
satisfied i.e
\be
e^x_{*_{RL}}\Box e^y_{*_{RL}}=e^{(x+y+z_2+z_3+\cdots)(\frac{R+L}{2})}
_{*_{RL}}\ ,
\ee
where
\be
z_2&=&\frac{1}{2}[x,y]_\Box\\
\nonumber\\
z_2&=&\frac{1}{12}[x,[x,y]_\Box]_\Box+\frac{1}{12}[[x,y]_\Box,y]_\Box
\hspace{1.0cm}{\rm etc.}
\ee
with the symbol (not a new $*$-commutator!),
\be
[x,y]_\Box&\equiv& x\Box y-y\Box x\equiv\frac{1}{2}x(*_{R}+*_L)y-\frac{1}{2}y
(*_R+*_L)x\nonumber\\
&\equiv& x\biggl(\frac{R+L}{2}\biggr) y-y\biggl(\frac{R+L}{2}\biggr)x\ .
\ee
We notice that if $[x,y]_\Box=0$, which means
\be
[x,y]_{*_{RL}}=0=[x,y]_{*_{LR}}\ ,\nonumber
\ee
then
\be
e^x_{*_{RL}}\Box e^y_{*_{RL}}=e^{x+y}_{*_{RL}}
\ee
We finally introduce the $*_{RL}$-coherent states\ ,
\be
|\alpha)&:=&e^{\alpha a^+}_{*_{RL}}|0>=\biggl(\frac{R+L}{2}\biggr)^{-1}
e^{(\frac{R+L}{2})\alpha a^+}|0>\nonumber\\
&=&e^{\alpha a^+(\frac{R+L}{2})}\biggl(\frac{R+L}{2}\biggr)^{-1}|0>
\ee
where $|0>$ stands for the vacum Fock state.
Calling $\frac{R(N)+L(N)}{2}\equiv f(N)$, and utilizing the number state
$|n>$, on which $f(N)|n>=f(n)|n>$ we obtain the expansion of the coherent state
in the Fock states,
\be
|\alpha)=f^{-1}(0)\biggl\{|0>+\frac{\alpha}{1!}f(0)|1>+\frac{\alpha^2}
{\sqrt{2!}}f(1)f(0)|2>+
\frac{\alpha^3}{\sqrt{3!}}f(2)f(1)f(0)\ |3>+\cdots\biggr\}
\ee
The normalized $*_{RL}$-CS is taken to be
\be
|\alpha>:=\frac{1}{\sqrt{(\alpha|\alpha)}}\ |\alpha)
\ee
where the overlap which provides the normalization factor reads,
\be
(\alpha|\alpha)=f^{-2}(0)+\frac{|\alpha|^2}{1!}+\frac{(|\alpha|^2)^2}{2!}
f(1)^2+\frac{(|\alpha|^2)^3}{3!}(f(2)f(1))^2+\cdots\ .
\ee
It is straightforward to prove that this coherent state is eigenstate of the
annihilation operator in the sence that,
\be
f^{-1}(N)af(N)\ |\alpha>=\alpha|\alpha>
\ee
or that
\be
a\cdot\biggl(\frac{R(N)+L(N)}{2}\biggr)\ |\alpha>=\alpha\biggl(\frac{R(N)+L(N)}
{2}\biggr)\ |\alpha>
\ee
which with the abbreviation issued by eq. (59) becomes,
\vskip 0.5 cm
\be
a\Box|\alpha>=\alpha\Box|\alpha>.
\ee
This eigenvalue problem is satisfied for any choise of the $*_{R,L}$-products
i.e. for any $q$-oscillator.
Before closing some remarks are in order. Despite the
demonstrated generality of the $*$-product approach to the deformed oscillator
we should note that all the particular choises above have the $*_R$ and $*_L$
products unequal, namely $R(N)\neq L(N)$ and they are both depended on $N$.
Open
 are
left a large class of deformed products (new $q$-oscillator) where $*_R$ and
$*_L$ are symmetric (i.e. $R(N)=L(N))$. Also cases of asymmetric products
such as $R(N)={\bf 1}$ and $L(N)\neq{\bf 1}$ are challenging, or cases
where one (or both) of the $*$-products are $c$-numbers (e.g. $R(N)=f(q)$
and $L(N)=f^{-1}(q))$ etc. are also interesting to explore. The
selection of the $*$-product is dictated only from the physical
relevance  of the ensuing $q$-deformed oscillator.

Finally, concerning future developments of the formalism put forward here
it would be interested to look at the classical level and search for a
classical version of the $*_{RL}$ -deformation of the commutator.
This will require
in addition to the ${\hbar }$ -deformation of the Poisson bracket, expressed
by the Moyal bracket, a new $q$ -deformation so that a two-parameter
deformation
scheme should be put forward [28].We expect to be able to formulate that
problem
using the r-matrix theory of the Lie-Poisson algebras[29] and we aim to return
to these matters elsewhere.
\vskip 1.0cm

\pagebreak


\begin{thebibliography}{99}
\bibitem{27}
This deformation of the dynamical equation is similar to that suggested by
ref.22, it would be interested to investigate further the relationships
between the
present formalism with that put forward in ref.22.
\end{thebibliography}
\end{document}